\documentclass[prl,aps,showpacs,groupedaddress,superscriptaddress,twocolumn,toc=flat]{revtex4-1}

\usepackage[colorlinks=true,linkcolor=blue,urlcolor=blue,citecolor=blue]{hyperref}

\usepackage[utf8x]{inputenc}
\usepackage{color}
\usepackage{bbm} 

\usepackage{amsfonts,amsmath,amssymb,stmaryrd}

\usepackage{graphicx}
\usepackage{subfigure}  
\usepackage{bbm} 
\usepackage{hyperref}
\usepackage{epsfig}
\usepackage{mathrsfs}
\usepackage{verbatim}
\usepackage{centernot}
\usepackage{ulem}
\usepackage{xspace}
\usepackage[english, status=draft]{fixme}       
\fxusetheme{color}                              %

\usepackage{float}

\newcommand{\bra}[1]{\langle#1|}
\newcommand{\ket}[1]{|#1\rangle}

\renewcommand{\H}{\hat{\mathcal{H}}}

\newcommand{\hc}{\text{h.c.}}

\newcommand{\tr}{\text{tr}}

\usepackage{array}

\usepackage{cancel,ifthen}
\newcommand{\cmnt}[2][NoInPuT]{\ifthenelse{\equal{#1}{NoInPuT}}{}{{\color{red}\sout{#1}}} {\color{blue} #2}}

\usepackage{bm}	

\begin{document}
\normalem	

\title{Analyzing non-equilibrium quantum states through snapshots\\ with artificial neural networks}

\author{A. Bohrdt}
\address{Department of Physics and Institute for Advanced Study, Technical University of Munich, 85748 Garching, Germany}
\address{Munich Center for Quantum Science and Technology (MCQST), Schellingstr. 4, D-80799 M{\"u}nchen, Germany}
\address{ITAMP, Harvard-Smithsonian Center for Astrophysics, Cambridge, MA 02138, USA}
\affiliation{Department of Physics, Harvard University, Cambridge, Massachusetts 02138, USA}

\author{S. Kim}
\affiliation{Department of Physics, Harvard University, Cambridge, Massachusetts 02138, USA}

\author{A. Lukin}
\affiliation{Department of Physics, Harvard University, Cambridge, Massachusetts 02138, USA}

\author{M. Rispoli}
\affiliation{Department of Physics, Harvard University, Cambridge, Massachusetts 02138, USA}

\author{R. Schittko}
\affiliation{Department of Physics, Harvard University, Cambridge, Massachusetts 02138, USA}

\author{M. Knap}
\address{Department of Physics and Institute for Advanced Study, Technical University of Munich, 85748 Garching, Germany}
\address{Munich Center for Quantum Science and Technology (MCQST), Schellingstr. 4, D-80799 M{\"u}nchen, Germany}

\author{M. Greiner}
\affiliation{Department of Physics, Harvard University, Cambridge, Massachusetts 02138, USA}

\author{J. L\'eonard}
\affiliation{Department of Physics, Harvard University, Cambridge, Massachusetts 02138, USA}


\date{\today}

\begin{abstract}
Current quantum simulation experiments are starting to explore non-equilibrium many-body dynamics in previously inaccessible regimes in terms of system sizes and time scales. Therefore, the question emerges which observables are best suited to study the dynamics in such quantum many-body systems. Using machine learning techniques, we investigate the dynamics and in particular the thermalization behavior of an interacting quantum system which undergoes a non-equilibrium phase transition from an ergodic to a many-body localized phase. 
We employ supervised and unsupervised training methods to distinguish non-equilibrium from equilibrium data, using the network performance as a probe for the thermalization behavior of the system. 
We test our methods with experimental snapshots of ultracold atoms taken with a quantum gas microscope. Our results provide a path to analyze highly-entangled large-scale quantum states for system sizes where numerical calculations of conventional observables become challenging.
\end{abstract}

\maketitle

\emph{Introduction.--}
After a global quench in a thermalizing system, local observables approach a value which corresponds to their expectation value in a typical microcanonical many-body eigenstate of the system~\cite{Deutsch1991,srednicki_chaos_1994,rigol_thermalization_2008}. 
Depending on the properties of the system and the initial state, the path to thermal equilibrium can vary. For example, conserved quantities can slow down the equilibration process \cite{Lux2014,Mukerjee2006,Bohrdt2017} or a quasi-stationary prethermal state can form, which exhibits properties different from the true thermal equilibrium state \cite{Berges2004}. 
\\
Quantum simulation experiments can enable the observation of the time-evolution of a quantum many-body system starting from a non-equilibrium state with almost perfect isolation from the environment. In the past decade, a variety of non-equilibrium phenomena has been observed with examples ranging from exotic phases realized through Floquet driving \cite{Aidelsburger2013,Miyake2013,Jotzu2014} to many-body localization \cite{Schreiber2015} and prethermalization \cite{Gring2012}.
\\
In many cases, theory can provide a clear prediction which observables should be studied, such as a given order parameter for a well-known phase transition. For some problems, however, it is not as clear which observable to look at, and by making a choice for one specific quantity, valuable information might be discarded. 
In many platforms with microscopic readout, Fock space snapshots of the quantum many-body state are the measured data set.
Fock space snapshots provide a wealth of information about the quantum many-body state by providing access to both local observables and non-local, high-order correlations. 
\\
\begin{figure}[h!]
\epsfig{file=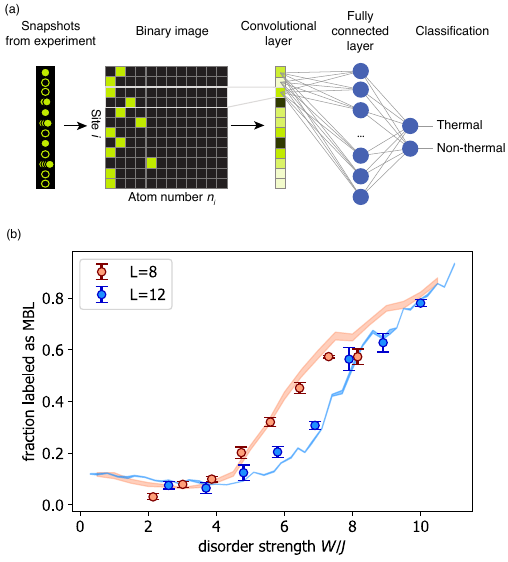, width=0.49\textwidth}
\caption{\textbf{Machine learning many-body localization.} The Bose-Hubbard model with a quasi-periodic disorder potential exhibits a many-body localized (MBL) phase, where thermalization breaks down, as the disorder strength is increased beyond a critical value. a) We study the dynamics of the system after a quench for different disorder strengths by evaluating snapshots from a quantum gas microscope with neural networks.
b) A neural network is trained to distinguish exact diagonalization snapshots at $W/J=0.3$ and $W/J=11$ for $U/J=2.9$ and a system with $8$ and $12$ sites at time $tJ=100$ after a global quench. After the training process is finished, snapshots at intermediate values of the disorder strength are used as input. The plot shows the resulting classification for numerical data (shaded band) as well as experimental snapshots (symbols). As the system size is increased, the fraction of snapshots classified as MBL begins to increase at larger values of $W$, indicating the transition in the finite size system.
The accuracies are averaged over two independent runs and the errors denote one s.e.m. 
}
\label{fig:schema}
\end{figure}
In order to address the challenge of finding suitable observables, artificial neural networks have recently emerged as a valuable tool in quantum many-body physics~\cite{Torlai2018,Carleo2019,Rem2018,Zhang2019,Bohrdt2019_ML}, and in nonequilibrium statistical mechanics \cite{Seif2020}. Previous machine learning approaches to study non-equilibrium systems have focused on quantities such as the entanglement spectrum~\cite{Schindler2017,Venderley2018,Hsu2018} or full eigenstates \cite{Zhang2019_mbl}, which are, however, experimentally inaccessible.
\\
In this work we study the dynamics of an interacting quantum many-body system in terms of experimental Fock space snapshots with the help of neural networks, Fig.~\ref{fig:schema}a). 
We find this analysis to have two main advantages: \emph{(i)} these snapshots are directly measured in many quantum simulation platforms, and large numbers of snapshots can be routinely obtained. \emph{(ii)} Raw data is used, 
where no analysis for specific quantities has taken place and all available information can be used without any bias. 
We consider the one-dimensional Bose-Hubbard model
\begin{equation}
\H = \sum_i \left[ -J \left(\hat{a}_i^\dag \hat a_{i+1} +\hc \right) + \frac{U}{2}  \hat n_i (\hat n_i - 1)
+W h_i \hat n_i \right].
\label{eq:BoseHubbard}
\end{equation}
Here, $\hat a^{(\dag)}_i$ annihilates (creates) a boson on site $i$ and $\hat n_i = \hat a^\dag_i \hat a_i$ is the particle number operator. The first term corresponds to hopping between neighboring sites, the second term is the interaction, here fixed at $U/J=2.9$, and the last term is the quasi-periodic potential mimicking on-site disorder with amplitude $W$, which can be created in a cold atom setup with an incommensurate lattice as $h_i = \cos (2\pi \beta i + \phi)$. In this work, we consider  $1/\beta=1.618$.
\\
This system exhibits a many-body localized (MBL) phase, where thermalization breaks down as the disorder strength is increased beyond a critical value. The transition from an ergodic to a many-body localized phase is fundamentally different from the well-studied case of equilibrium phase transitions, as it describes a non-equilibrium setting~\cite{Basko2006,Gornyi2005,Oganesyan2007,Pal2010,Serbyn2013,Huse2014,Serbyn2014,Abanin2018,Chiaro2020}. 
Finding the transition point is numerically challenging, because it is usually obtained from entanglement properties or the level statistics, which can only be obtained for small system sizes where full diagonalization of the Hamiltonian is possible.
Here, we focus on Fock space snapshots of the many-body quantum state as input data, which are the direct output of quantum gas microscopy experiments and thus experimentally readily accessible for the systems of interest. This approach has the advantage that significantly bigger system sizes can be reached experimentally. 
\\
We consider the dynamics of two one-dimensional systems of $8$ and $12$ sites, which are initialized in a Mott-insulating state with exactly one particle per site.
In Fig.~\ref{fig:schema}, we first train the network to distinguish snapshots of the many-body quantum state, obtained from exact diagonalization calculations, for low ($W/J=0.3$) and high ($W/J=11.0$) disorder strength for an interaction strength of $U/J=2.9$ in the comparatively long-time limit at time $tJ=100$.
We average over ten different disorder realizations, obtained by varying the phase $\phi$ in the potential. 
After the network has learned to label the extremal cases correctly with sufficiently high accuracy ($>90\%$), we input snapshots for intermediate values of the disorder strength. After training the neural network on numerically simulated snapshots, we use experimental data as input, where each snapshot stems from a different disorder realization. As output, for each disorder strength we obtain the fraction of snapshots labeled as \emph{many-body localized} and \emph{thermalizing}, see Fig.~\ref{fig:schema}. 
Based on these results, we conclude that the many-body localization transition is located within the range of $W/J \approx 4-8$ with strong finite-size drifts. This result is in agreement with previous experiments~\cite{Lukin2019,Rispoli2019}, which considered conventional observables such as the local entropy. Notably, the local entropy exhibits volume law scaling both in the thermal and the MBL phase and is thus by itself not sufficient to locate the transition without exact numerics~\cite{Lukin2019}. Our results, in contrast, are able to distinguish the two phases without any theoretical input, which suggests that the network learned a more suitable observable to distinguish the two phases. In \cite{supp}, we show the level statistics for system sizes $L=6,7,8$ for comparison.  Similar to the machine learning analysis of a disordered spin chain based on the entanglement spectrum in \cite{Schindler2017}, the transition found by the neural network is as sharp as the level statistics, but exhibits a small shift to larger disorder strengths. 
\begin{figure*}
\centering
\epsfig{file=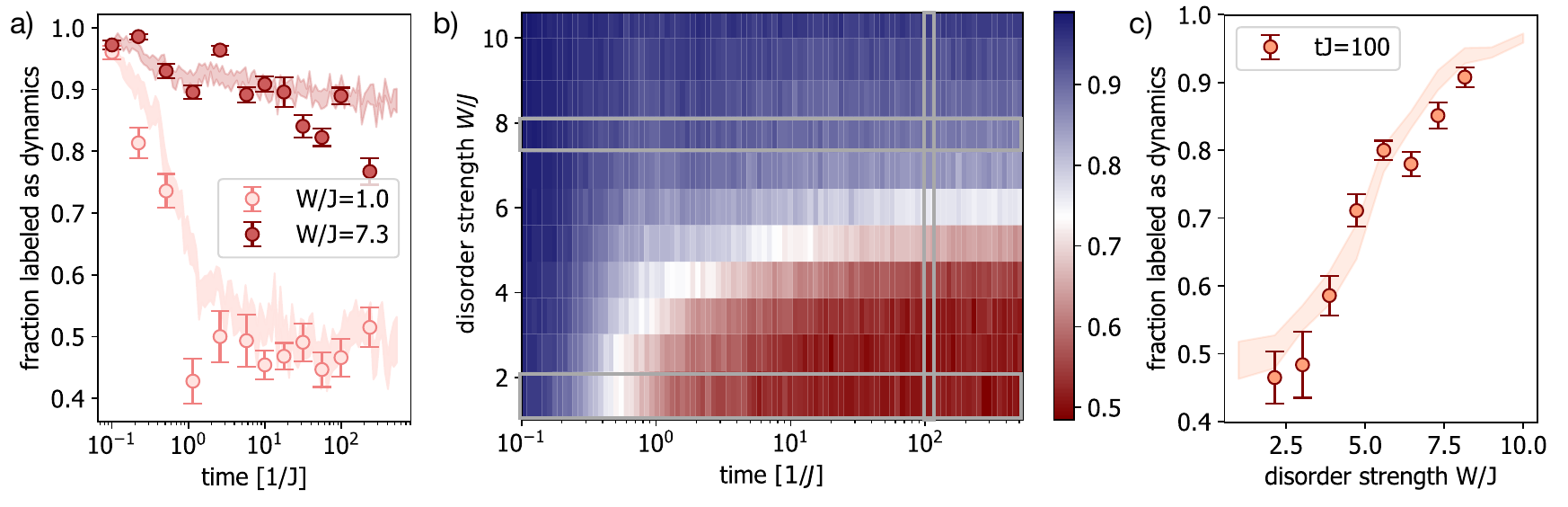, width=0.95\textwidth}
\caption{\textbf{Learning thermalization.} A system with $8$ sites and $U/J=2.9$ is initialized in a Mott-insulating state of one particle per site and the ensuing time evolution is investigated. In each time step, the neural network is trained to distinguish snapshots from the current time step from snapshots from a thermal state with the same energy density, both obtained from exact diagonalization. A high accuracy indicates that the current time step can be easily distinguished from the thermal state. 
a) The resulting classification as \emph{dynamics} versus \emph{equilibrium} for $W/J=1.0$ and $W/J=7.3$, averaged over $12$ different disorder realizations (shaded line). Experimental data from the dynamics after the quench is used as input at selected time steps (symbols). 
b) Exact diagonalization results for disorder strengths between $W/J=1$ and $W/J=10$ for the full dynamics.
c) Classification as \emph{dynamics} versus \emph{equilibrium} at time $tJ=100$ for disorder strengths between  $W/J=1$ and  $W/J=10$. The results are averaged over $10$ independent runs and the error bars correspond to the s.e.m. 
}
\label{fig:thermalization}
\end{figure*}

While we have only compared two extremal disorder strengths in the long-time limit, the full dynamics of the system contain  much more information. We proceed by analyzing the time- and disorder-strength dependence of the system after the global quench.
\\
\\
\emph{Learning thermalization.--}
We now investigate the system's approach to thermal equilibrium by comparing each time step to a thermal state of the same Hamiltonian. The performance of the network in distinguishing dynamics from equilibrium can then be used as a probe of thermalization.
\\
In order to compare the time evolved state to thermal equilibrium, all conserved quantities of the model should be considered~\cite{rigol_thermalization_2008}. 
In our experiment, both the energy density and the particle number are conserved during the many-body evolution. 
The energy density of the initial state is matched by choosing the temperature of the thermal state accordingly. 
We take the conservation of the total particle number into account by calculating the thermal state within a fixed particle number sector.
We numerically generate snapshots from such a state in thermal equilibrium as well as from the time-evolved state for each time step under consideration.
\\
For each time step, we train the network to label the snapshots from the thermal equilibrium distribution as \emph{equilibrium}, and the snapshots from the numerically time-evolved initial state as \emph{dynamics}. The neural network parameters optimized for each time step seperately.
We then test the network's performance by inputting experimental data with different evolution times.
In Fig.~\ref{fig:thermalization}a) the resulting classification into the categories \emph{dynamics} versus \emph{equilibrium} is shown as a function of time. Here, we average over $12$ different disorder realizations and take snapshots at the corresponding effective temperatures. 
\\
For small $W/J$, the system thermalizes comparably fast: for times $tJ > 10$, the network reaches an accuracy of $50\%$, equivalent to guessing between the two classes. This means the network fails to distinguish snapshots from the time-evolved state from the corresponding thermal state.
For high values of $W/J$ the system fails to thermalize on the time-scales accessed here, and the network is able to distinguish the current timestep from the thermal equilibrium state with a high accuracy.  Using an interpretable network architecture \cite{Miles2020}, we find that for intermediate disorder strengths, higher order correlations play a role in the classification task, see \cite{supp}. 
\\
We study the long time limit at $tJ=100$ for a range of values of the disorder strength. 
As shown in Fig.~\ref{fig:thermalization}c), the fraction of snapshots classified as \emph{dynamics} rises strongly between $W/J\approx 4$ and $W/J\approx 8$ and reaches values close to $1$, indicating that the system has not reached thermal equilibrium.
\\
We benchmark our experimental results by testing the network with theoretical snapshots not used during training and find good agreement throughout the range of the covered parameters.
\\
This procedure has the advantage that the features used to make the classification can vary for different time steps and the network specifically searches for differences between the current time and thermal equilibrium. It is therefore in principle capable of identifying specific observables that have not yet reached their thermal equilibrium value and thus find, for example, (almost-) conserved quantities. 
Indeed, with this method we find deviations from thermal equilibrium already in the range of $W/J \approx 2-5$, in contrast to the results from the classification scheme in Fig.~\ref{fig:schema}b). This indicates an improved sensitivity of our method. 
Here we consider a system which exhibits a transition from thermalizing behavior to many-body localization, which constitutes a canonical example in the study of non-equilibrium phenomena. Note, however, that our scheme is not limited to the system considered here and can be applied to a variety of models. This method also allows to detect, for example, prethermal behavior and the existence of conserved quantities that keep their value during the dynamics and therefore never reach a generic thermal equilibrium value.  Another canonical model to study equilibration behavior is the transverse field Ising model,  which has an extensive number of conserved quantities. In \cite{supp}, we show that a neural network performs significantly worse in distinguishing the time-evolved state from an approximative generalized Gibbs ensemble, where a few conserved quantities are taken into account, than the simple thermal state discussed above, where only the energy density is considered.  This highlights the capability of our approach to identify conserved quantities, which can drastically alter the thermalization process.  
Our method comes at the expense that one needs snapshots from the thermal density matrix for training, which -- especially in the case of a non-thermalizing phase such as MBL -- may need to be generated numerically. 
In the following, we overcome this limitation by analyzing the transition in the dynamics with an unsupervised scheme that, in principle, does not rely on theory data.
\\
\\
\emph{Confusion learning.--}
\begin{figure}
\centering
\epsfig{file=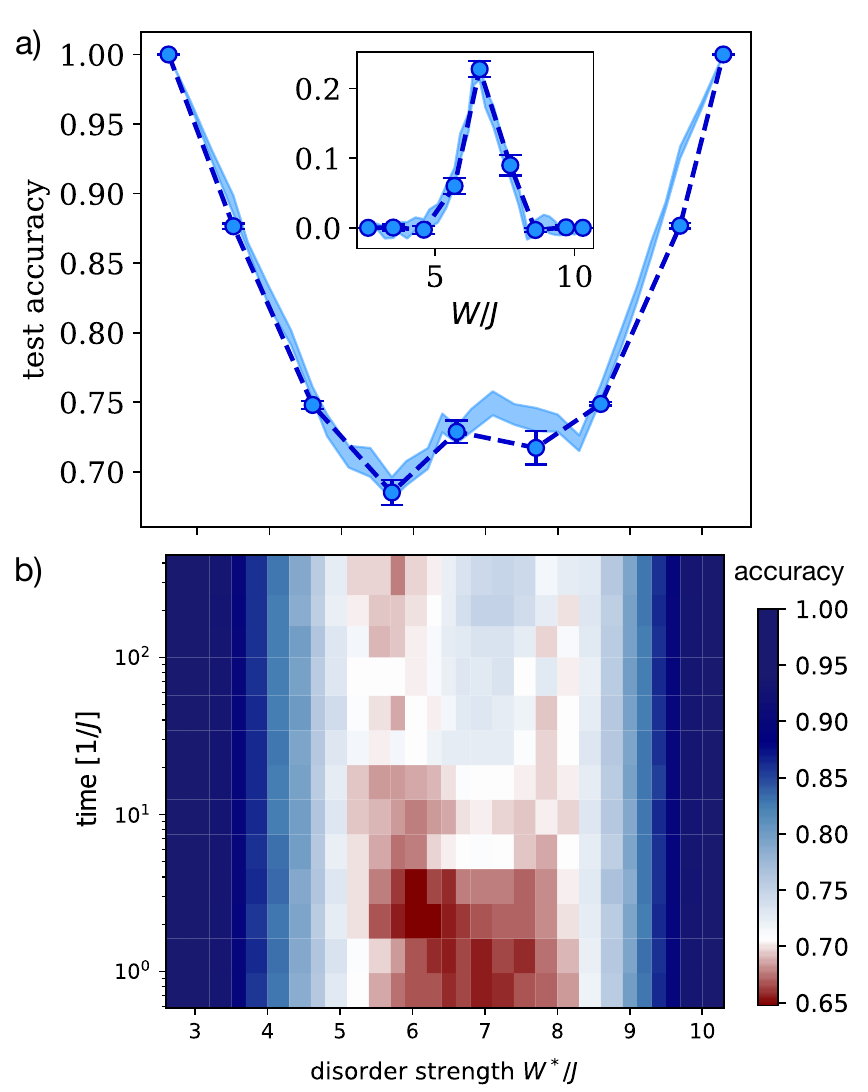, width=0.5\textwidth}
\caption{\textbf{Confusion learning.} Snapshots of the many-body quantum state of a system with $12$ sites, $U/J=2.9$, and various disorder strengths $W/J$ are analyzed using the confusion learning scheme. A neural network is trained to label all snapshots with $W<W^*$ as \emph{phase A} and the remainder as \emph{phase B}. If a qualitative change in the data occurs, the accuracy will peak at an intermediate value of $W^*$. a) The resulting accuracy at time $tJ=100$ after the global quench for training on numerically simulated data (shaded line) and sorting experimental data (symbols). Inset: same data after subtracting the accuracy for randomly labeled data.
b) The accuracy for repeating the training process for different time points during the dynamics after the quench using numerically simulated data. The results are averaged over $10$ independent runs and the error bars correspond to the error based on one s.e.m. 
}
\label{fig:unsupervised}
\end{figure}
Several unsupervised learning schemes that use the network performance to probe whether and where a phase transition or more general, a qualitative change in the data, exists have been proposed \cite{Nieuwenburg2017,Schaefer2019,Greplova2020}.
Here, we adapt a scheme termed ``confusion learning'' introduced in Ref.~\cite{Nieuwenburg2017}.
In brief, the scheme works as follows: We have a dataset of snapshots for values of the disorder strength $0.3 \leq W/J\leq 11.0$. 
The goal is to test whether a value $W^*$ exists at which the data changes qualitatively. We start with a guess for $W^*$ and label all snapshots for $W\leq W^*$ as \emph{phase A} and correspondingly all snapshots with $W>W^*$ as \emph{phase B}. Assuming the snapshots are qualitatively different for $W\leq W^*$ as compared to $W>W^*$, the network should achieve a high accuracy in assigning the correct labels. However, if there is no qualitative change at the $W^*$ under consideration, there will be confusion about the correct labels and the accuracy will thus be lower. Therefore, if there is a qualitative change in the data, the accuracy as a function of $W^*$ will be maximal if $W^*$ corresponds to the transition point. Trivially, the test accuracy is expected to approach unity when the guessed $W^*$ corresponds to the minimum or maximum value of $W$, because all data are labelled equally and no confusion occurs. In total, the presence of a critical point is therefore signalled by a characteristic \emph{W}-shape of the test accuracy as a function of the control parameter.
\\
We train the neural network with numerical snapshots in the long-time limit ($tJ=100$) in order to test for the presence of a phase transition. Subsequently, we use experimental data as input to the network, Fig.~\ref{fig:unsupervised}a). The data shows the onset of a maximum around $W^*/J=7$, indicating the presence of a critical point in agreement with Fig.~\ref{fig:schema}b).  The contrast in the \emph{W}-shape achieved here is comparable to the signal seen for a spin model in \cite{Nieuwenburg2017}, where instead of snapshots the entanglement spectrum is used as input to the neural network.  
In order to isolate the signal of the phase transition from the trivial part of the \emph{W}-shape, we subtract the accuracy obtained when training on randomly labeled data. The resulting difference, shown in the inset of Fig.~\ref{fig:unsupervised}a), exhibits a clear peak at $W^*/J=7$, that indicates the transition between the different dynamical phases.
We also check with theoretical snapshots not used during training and find qualitatively similar behaviour. We attribute the slight deviation in the maximum to the coarse resolution in the disorder strength for the experimental data.
\\
Since differences in the thermalization behavior only present themselves in the course of the dynamics, we expect the phase transition to remain hidden at short evolution times. In order to reveal this effect, we perform the same method with theoretical snapshots at different evolution times. In Fig.~\ref{fig:unsupervised}b), the resulting accuracy achieved by the network is shown as a function of $W^*$. 
\\
These results have several advantages compared to the previous methods: as opposed to Fig.~\ref{fig:schema}b), we do not a priori assume that there is a transition. Moreover, we specifically train the network to find differences between the snapshots at all available values of the disorder strength, thus avoiding bias from the choice of training data.
\\
\\
\emph{Summary and Outlook.--}
In this work, we used machine learning techniques to study the non-equilibrium dynamics after a global quench in the one-dimensional Bose-Hubbard model with a quasi-periodic disorder potential. 
We used supervised as well as unsupervised machine learning methods to probe for a qualitative change in experimental snapshots as the disorder strength is tuned. 
Comparing the results for systems with $8$ and $12$ sites, we find that the critical value of the disorder strength increases with the system size, proving the need for methods applicable in large -- experimentally accessible -- systems. In contrast to standard tools to locate the MBL transition, the methods used here can be directly applied to experimental data taken with a quantum gas microscope and are not limited to small system sizes. 
We furthermore studied the approach to thermal equilibrium -- or lack thereof -- by training a neural network to distinguish snapshots from the current time step from snapshots from a thermal ensemble at the same energy and particle density. The accuracy achieved by the network indicates how non-thermal the time-dependent quantum many-body state is. 
\\
An exciting future research direction consists of applying the same scheme to identify conserved or almost-conserved quantities in experimentally accessible data, for example by using a generalized Gibbs ensemble for comparison.
 Apart from the concrete system studied here, it would be interesting to consider other models and phenomena, for example quantum scars \cite{Bernien2017,Turner2018} and Hilbert space fragmentation~\cite{Sala2020,Khemani2020,Scherg2020}. In order to gain additional physical insights, interpretability is an extremely important direction for future work and it would be interesting to study which observables the network uses to make the classifications considered here \cite{Miles2020}, and how those observables change during the time evolution of the many-body system. 
\\
\\
\emph{Acknowledgements.--}
We would like to thank Eugene Demler, Fabian Grusdt, Florian Kotthoff,  Cole Miles, and Frank Pollmann for fruitful discussions. 
We acknowledge support from the Technical University of Munich - Institute for Advanced Study, funded by the German Excellence Initiative and the European Union FP7 under grant agreement 291763, the Deutsche Forschungsgemeinschaft (DFG, German Research Foundation) under Germanys Excellence Strategy--EXC--2111--390814868, TRR80, DFG grant No. KN1254/2-1, No. KN1254/1-2, from the European Research Council (ERC) under the European Unions Horizon 2020 research and innovation programme (grant agreement No. 851161), from the NSF Graduate Research Fellowship Program (S.K.), 
from the NSF grant PHY-1734011, NSF grant PHY-1806604l, NSF grant OAC-1934598, the Gordon and Betty Moore Foundations EPiQS Initiative, the Vannevar Bush Award, and the Swiss National Science Foundation (J. L.).

\clearpage
\newpage
\section*{Supplementary information}

\section{Transition}

\begin{figure}[t!]
\centering
\epsfig{file=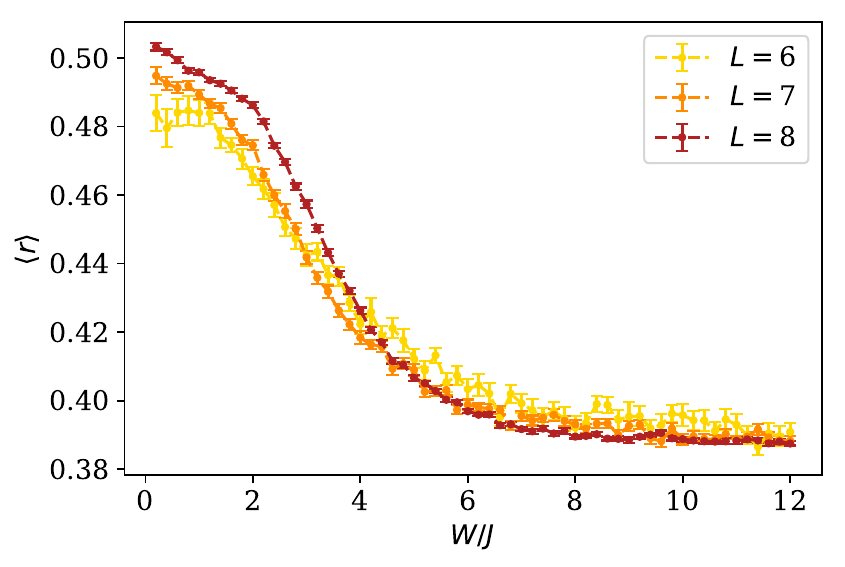, width=0.5\textwidth}
\caption{\textbf{Level statistics.} Average value of the ratio of adjacent energy gaps for different system sizes $L=6,7,8$ at a density of one particle per site as a function of disorder strength. } 
\label{fig:levelstats}
\end{figure}

Upon increasing the disorder strength, the level statistics of the Hamiltonian evolves from the Gaussian-orthogonal ensemble to Poisson statistics as the system enters the MBL phase \cite{Oganesyan2007,Pal2010}. We consider the level spacings
\begin{equation}
\delta_\phi^{(n)} = |E_\phi^{(n)} - E_\phi^{(n-1)}|,
\end{equation}
where $E_\phi^{(n)}$ is the $n$-th eigenenergy of Hamiltonian (1) in the main text with disorder realization given by $\phi$. The ratio of adjacent gaps is then given as
\begin{equation}
r_\phi^{(n)} = \text{min}\left(\delta_\phi^{(n)},\delta_\phi^{(n+1)} \right) / \text{max}\left(\delta_\phi^{(n)},\delta_\phi^{(n+1)} \right).
\end{equation}
For a given system size $L$, we fix the particle density to one particle per site and for each disorder strength $W/J$ obtain the average value of this ratio over $30$ disorder realizations, i.e. different values of $\phi$. In Fig.~\ref{fig:levelstats}, the resulting average value of the ratio of adjacent energy gaps $\left<r \right>$ is shown as a function of the disorder strength for system sizes $L=6,7,8$. The shift of the drop in $\left<r \right>$ to larger values of $W/J$ agrees well with the results presented in Fig.1b) of the main text.

\begin{figure}[t!]
\centering
\epsfig{file=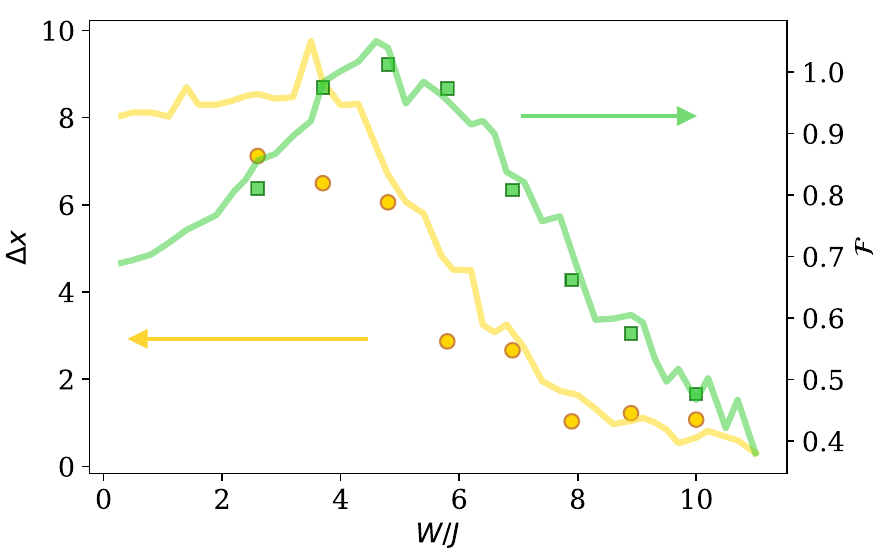, width=0.5\textwidth}
\caption{\textbf{Observables.} The transport distance $\Delta x$ and on-site fluctuations $\mathcal{F}$, see text, are evaluated from the same snapshots. Shaded bands correspond to exact diagonalization snapshots, symbols are based on experimental data. We simultaneously evaluate snapshots from ten different disorder realizations in the numerical data. In the experimental data, each snapshot is from a different disorder realization. The errors denote one s.e.m.  } 
\label{fig:observables}
\end{figure}

In order to relate to previous work, in particular Ref.~\cite{Rispoli2019}, we directly evaluate observables from the snapshots and calculate the transport distance $\Delta x$, defined as
\begin{equation}
\Delta x = 2\sum_d |d| \cdot \left<G_c^{(2)}(i,i+d)\right>_i
\label{eq:transportDeltaX}
\end{equation}
with 
\begin{equation}
G_c^{(2)}(i,i+d) = \left< \hat{n}_i \hat{n}_{i+d}\right> - \left<\hat{n}_i \right> \left< \hat{n}_{i+d}\right>,
\end{equation}
and the on-site fluctuations $\mathcal{F}$, defined as
\begin{equation}
\mathcal{F} = G_c^{(2)}(d=0),
\end{equation}
in Fig.~\ref{fig:observables}. Comparing the output of the neural network in Fig.1b) of the main text with the transport distance $\Delta x$  shows a similar behavior, from which one might conjecture that the network uses a similar observable to make the distinction. Note that with this approach, we are able to make a quantitative prediction on the basis of single or few snapshots, for which the observables shown in Fig.~\ref{fig:observables} are not clearly converged to their average value. 

\subsection{Confusion learning: experimental data}
\begin{figure}
\centering
\epsfig{file=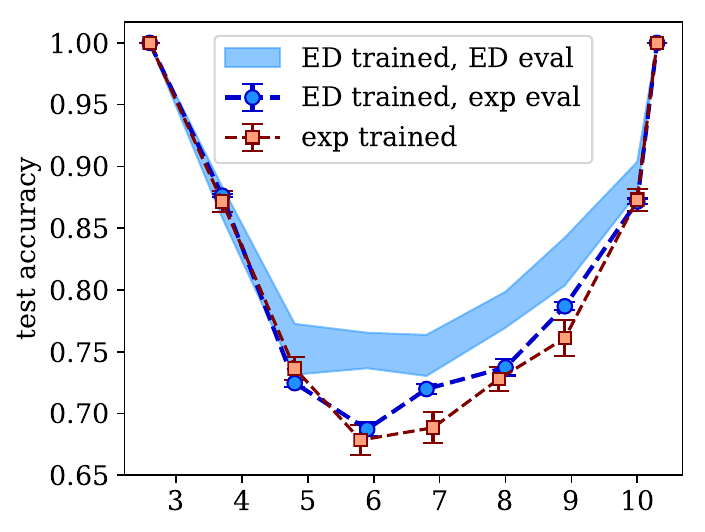, width=0.5\textwidth}
\caption{\textbf{Confusion learning - experimental data.} Same analysis as shown in Fig.3 of the main text with training on experimental data (red symbols).  The same analysis is performed for numerical data from exact diagonalization, using the same values of the disorder strength.  
}
\label{fig:confusion_exp}
\end{figure}
In Fig.~\ref{fig:confusion_exp}, we show the same analysis as presented in Fig.3 of the main text, but using a network that has been solely trained on experimental data.  A clear ``W''-shape does not emerge.  The experimental result agrees qualitatively with the result using exact numerical data, where we only used data at the same values of disorder strength as available from the experiment. 
We hence attribute the lack of a clear ``W''-shape to the rough grid of values of the disorder strength -- consisting of eight values between $W/J=2$ and $W/J=10$. 

\subsection{Unsupervised learning of the transition}

\begin{figure}[ht!]
\centering
\epsfig{file=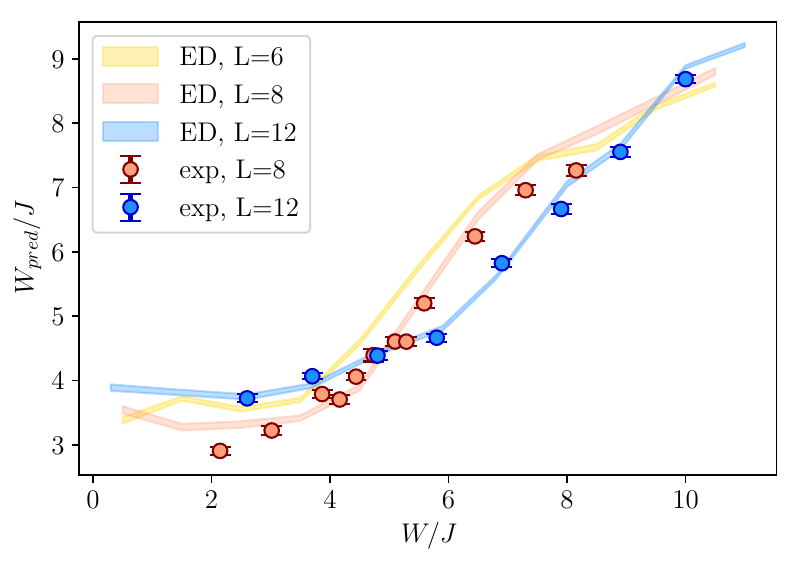, width=0.5\textwidth}
\caption{\textbf{Unsupervised learning of a phase transition.} Snapshots of the many-body quantum state of a system with $6$, $8$, and $12$ sites, $U/J=2.9$, and various disorder strengths $W/J$ are analyzed using the unsupervised learning scheme introduced in \cite{Schaefer2019,Greplova2020}. A neural network is trained to label a given snapshot with the corresponding value of $W/J$. If a qualitative change in the data occurs, the derivative $\delta W_{\text{pred}} / \delta W_{\text{label}}$ will exhibit a maximum. The plot shows $W_{\text{pred}}$ as a function of $W_{\text{label}}$. Shaded bands are exact diagonalization and symbols experimental data.
The results are averaged over $25$ independent runs and the error bars corresponds to one s.e.m. .} 
\label{fig:unsupervised_Wpred}
\end{figure}

We use the unsupervised scheme introduced in \cite{Schaefer2019,Greplova2020} to locate the transition to the many-body localized phase as a function of the disorder strength $W$. In Fig.~\ref{fig:unsupervised_Wpred}, we show the predicted values of the disorder strength as a function of the actual values, $W_{\text{label}}$. The experimental data agrees well with numerics. 
The steepest slope, indicating the transition, shows a similar shift as observed in the supervised learning scheme used in Fig.1 of the main text.

\section{Learning thermalization}

\begin{figure}
\epsfig{file=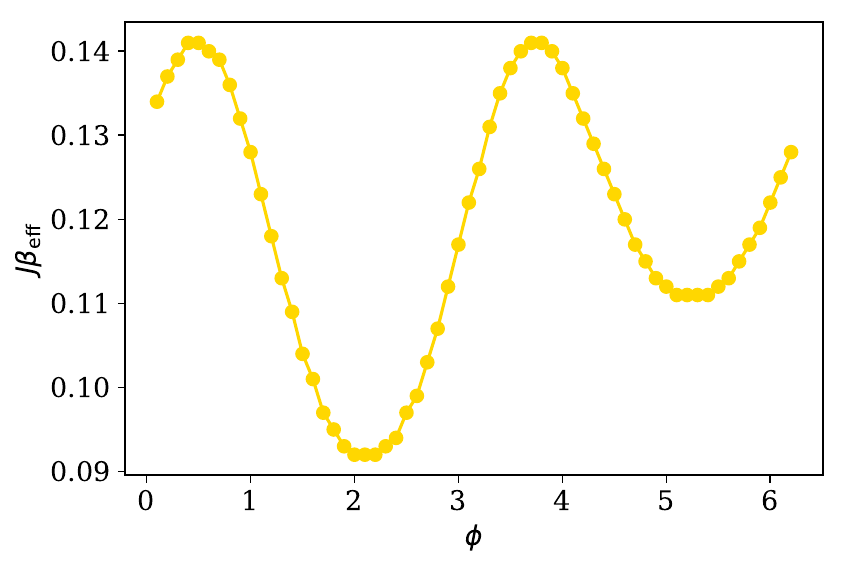, width=0.5\textwidth}
\caption{\textbf{Effective temperatures.} Effective inverse temperatures $\beta_{\rm{eff}}$ for $U/J=2.9$, $W/J=4.5$ as a function of $\phi$ for a system with $L=6$ sites and a density of one particle per site.} 
\label{fig:effectiveT}
\end{figure}

In order to compare the time evolved state to thermal equilibrium, all conserved quantities of the model should be taken into account. The energy density of the initial state can be matched by choosing the temperature of the thermal state accordingly. In particular, the energy density of the initial state $\ket{\psi_0}$ is given by $E_i = \bra{\psi_0} \hat{H} \ket{\psi_0}$.
The effective temperature $T_{\text{eff}}$ is then determined such that the density matrix of the system,
$\hat{\rho}_\beta= \frac{1}{Z} \exp(-\beta_{\text{eff}} \hat{H})$,
with the inverse temperature $\beta_{\text{eff}} = 1/T_{\text{eff}}$ and $Z = \tr ( \exp(-\beta_{\text{eff}} \hat{H}) )$ fufills
\begin{equation}
E_i = \tr \left( \hat{H} \hat{\rho}_\beta \right).
\label{eq:effectiveT}
\end{equation}
The energy density $E(\beta) = \tr ( \hat{H} \hat{\rho}_\beta )$ is calculated for a range of values $\beta$ until the effective temperature is determined such that Eq.~\ref{eq:effectiveT} is fulfilled. 
 Due to the disorder potential, this effective temperature varies for different values of $\phi$, where $\phi$ determines the disorder realization. In Fig.~\ref{fig:effectiveT}, the effective inverse temperature is shown as a function of $\phi$ for a system with $L=6$ sites at unity filling for interaction strength $U/J=2.9$ and disorder strength $W/J=4.5$.

\subsection{Interpretability: higher-order correlation functions}
\begin{figure*}
\centering
\epsfig{file=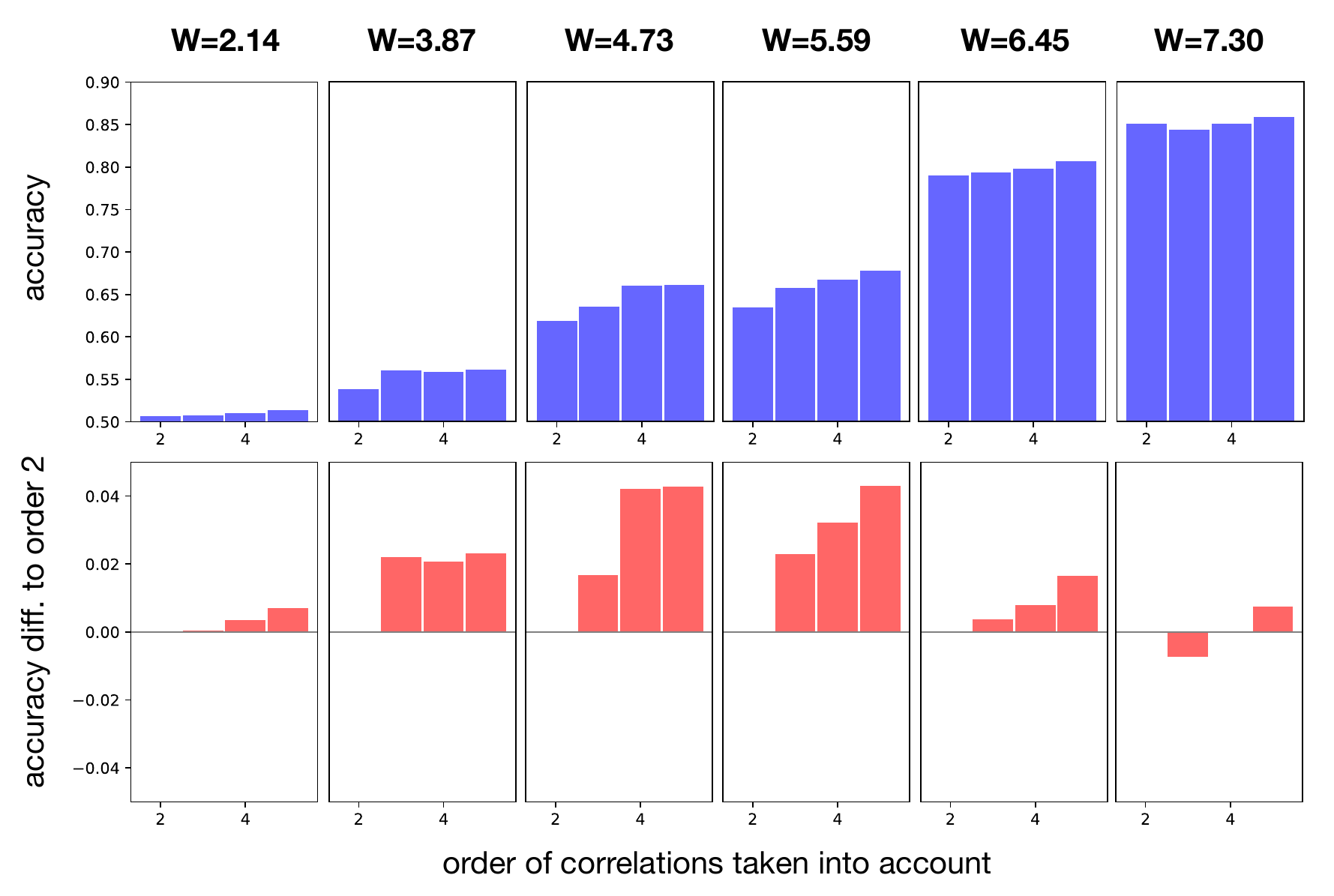, width=0.9\textwidth}
\caption{\textbf{Accuracy as a function of order of correlations.} Accuracy obtained for comparing snapshots from time step $tJ=7.3$ to a thermal ensemble using the correlator convolutional neural network (CCNN) architecture \cite{Miles2020}. The order of correlation functions used by the neural network serves as a hyper parameter in the CCNN architecture.  The top plot shows the obtained accuracy for different disorder strengths for correlation orders 2,3,4, and 5. The bottom plot shows the difference to the accuracy obtained when taking second order correlation functions into account to allow for a better comparison of the results between the different disorder strengths.}
\label{fig:order}
\end{figure*}

Following the same approach as in main text Fig. 2,  but using the correlator convolutional neural network (CCNN) architecture, we can gain insights into the information used to solve the classification task. In particular, the order of correlation considered enters as a hyper parameter of the network architecture. The network can thus be trained to distinguish dynamics from thermal equilibrium taking into account for example only correlations up to second order.  Note that given the Fock space snapshots we consider here, all correlations considered are density-density correlations.
In Fig.~\ref{fig:order},  we compare the accuracies obtained by the network when taking into account correlations up to second, third, fourth, and fifth order for different disorder strength, for comparing the time step $tJ=7.3$ to thermal equilibrium.  Since the overall scale increases significantly with increasing disorder strength, as the state becomes less and less thermal,  we show in Fig.~\ref{fig:order} bottom the accuracies obtained for correlations of order $3-5$ with the accuracy for order $2$ subtracted.  For low and high disorder,  taking into account correlation functions of order higher than two does not significantly increase the accuracy obtained by the CCNN.  However, for intermediate disorder strengths, higher order correlations play a role in the classification task and the accuracy improves by up to $5\%$ when considering higher order correlators. This result is in accordance with Ref.~\cite{Rispoli2019},  which showed sizable higher order correlations in the critical regime at intermediate disorder strength.

\subsection{Transverse field Ising model and generalized Gibbs ensemble}
\begin{figure}
\centering
\epsfig{file=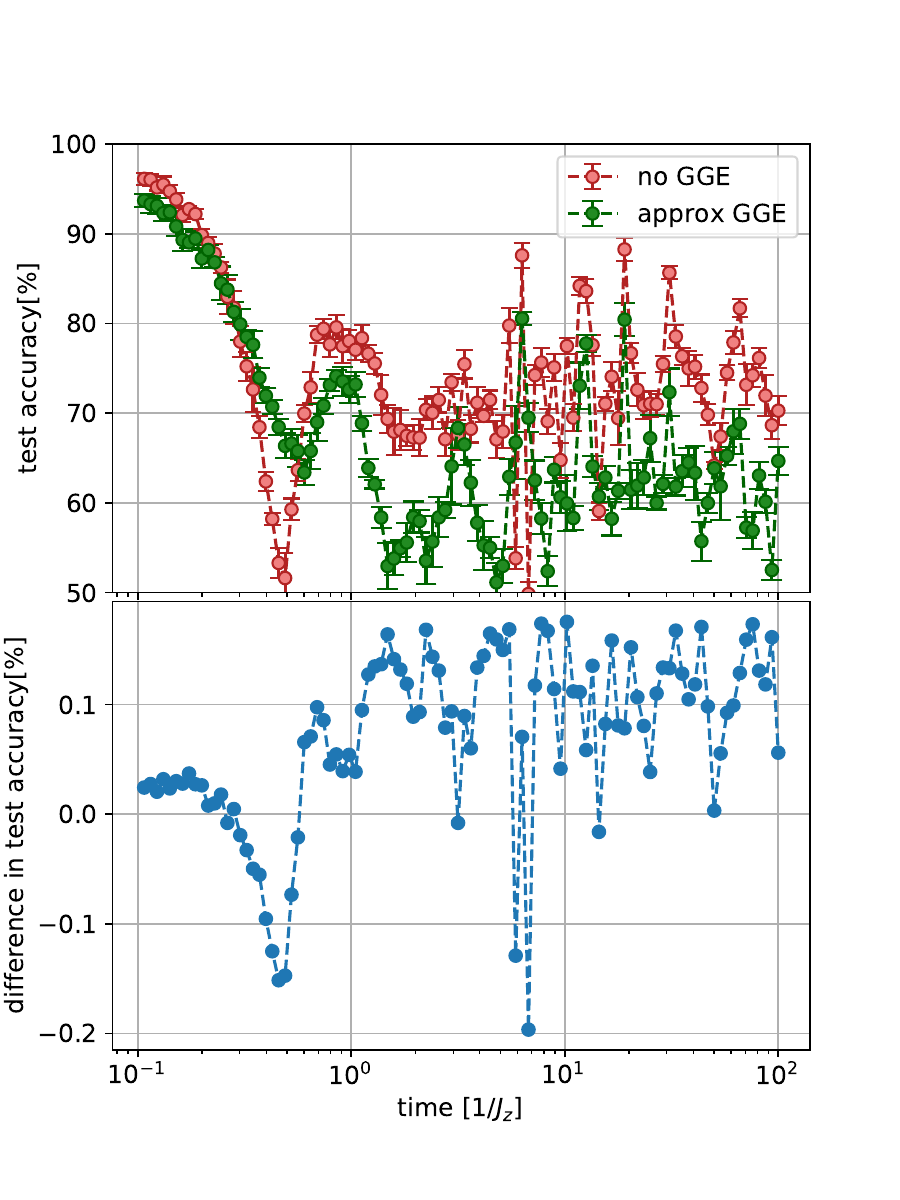, width=0.5\textwidth}
\caption{\textbf{Thermalization and conserved quantities.} We consider the dynamics after initializing the system in a product state in the transverse field Ising model (TFIM). In the TFIM, there is an extensive number of conserved quantities, which have to be considered to correctly describe the long-time limit.  Top: If only the conserved energy is taken into account (red), the long-time limit is still comparably easy to distinguish from the thermal density matrix. The performance of the network drops significantly if the first (i.e., most local) two conserved quantities are additionally taken into account in an approximation of the generalized Gibbs ensemble (green).  Bottom: Difference between the accuracies shown in the top plot.}
\label{fig:TFIM}
\end{figure}

The transverse field Ising model,
\begin{equation}
\hat{H} = -J \sum_i \left(\hat{\sigma}_i^z \hat{\sigma}_{i+1}^z + h \hat{\sigma}_i^x\right),
\end{equation}
has an extensive number of local conservation laws, which are known to be \cite{Grady1982,Prosen1998}
\begin{equation}
I^{(n,+)} = -J\sum_j (S_{j,j+n}^{xx} + S_{j,j+n+2}^{yy}) + h (S_{j,j+n-1}^{xx} +S_{j,j+n-1}^{yy})
\end{equation}
and
\begin{equation}
I^{(n,-} = -J\sum_j(S_{j,j+n}^{xy}-S_{j,j+n}^{yx})
\end{equation}
with $I^{(1,+)} = H$ and 
\begin{equation}
S_{j,j+l}^{\alpha \beta} = \sigma_j^\alpha \left[ \prod_{k=1}^{l-1} \sigma_{j+k}^z \right] \sigma_{j+l}^\beta, \qquad S_{j,j}^{yy} = -\sigma_j^z.
\end{equation}
We numerically simulate the dynamics for $J=1$, $h=2$ starting from the initial product state
\begin{equation}
\ket{\psi_0} = \prod_i \left( \sin \theta/2 \ket{\uparrow}_i \cos \theta/2 \ket{\downarrow}_i \right)
\end{equation}
with $\theta = 3.0$ using exact diagonalization for a system of size $L=12$.  We then train a neural network to distinguish snapshots sampled from the time-evolved state from a thermal ensemble. For the thermal ensemble, we consider two different choices:
\begin{itemize}
\item a thermal ensemble $\rho = 1/Z \exp(-\beta H)$, where the inverse temperature $\beta$ is determined by the energy $\bra{\psi_0} H \ket{\psi_0}$ as discussed above
\item an approximation to the generalized Gibbs ensemble, $\rho_{\text{GGE}} = 1/Z_{\text{GGE}} \exp(-\beta H -\lambda_2 I^{2,+} - \lambda_3 I^{3,+})$ with Lagrange multipliers $\lambda_{2,3}$ determined in the same way for the conserved quantities defined above, with $Z_{\text{GGE}}$ the corresponding partition sum for the approximation of the GGE, such that $\text{tr} \rho_{\text{GGE}} = 1$.  We determine the parameters to match the expectation values in the initial state, yielding $\beta=0.304$, $\lambda_2 = 0.336$, and $\lambda_3=0.1$.
\end{itemize}

As shown in Fig.~\ref{fig:TFIM}, the performance of the neural network in distinguishing dynamics from thermal/generalized Gibbs ensemble is significantly worse in the latter case, indicating that the long-time dynamics is better described by the GGE. This emphasizes the importance in taking conserved quantities into account in order to correctly describe the equilibration behavior, and in particular the ability of a neural network to capture the differences between the time evolved state and the standard thermal ensemble.

\subsection{Distinguish from long-time limit}
\begin{figure}
\centering
\epsfig{file=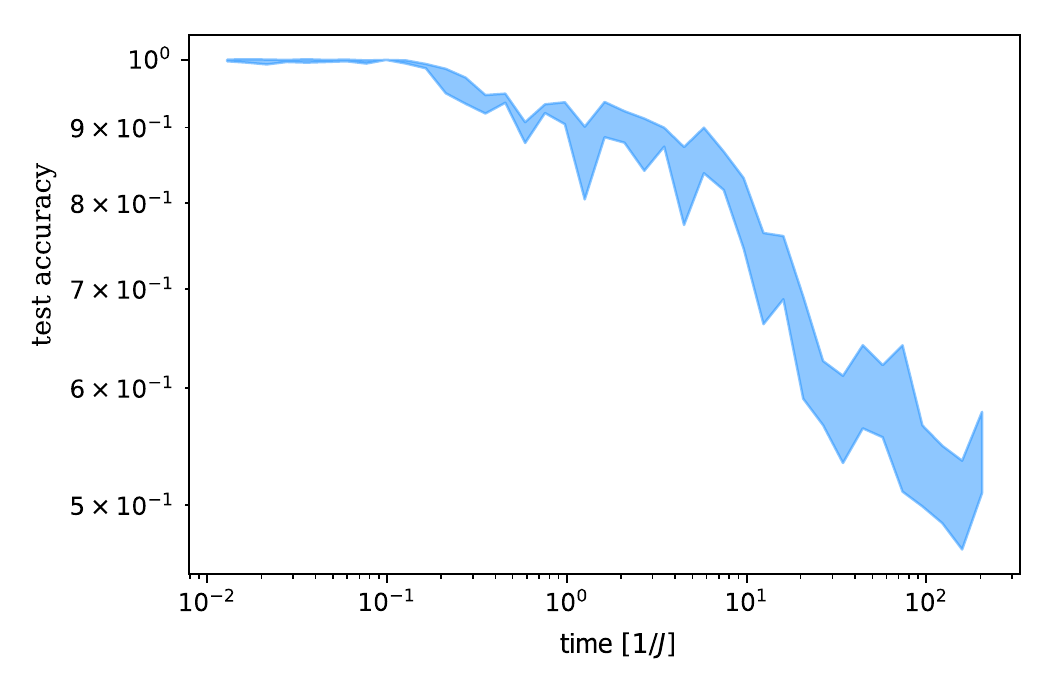, width=0.5\textwidth}
\caption{\textbf{Learning thermalization.} The system is initialized in a uniform state ($\ket{111111111111}$) and the ensuing time evolution is investigated. In each time step, the neural network is trained to distinguish snapshots from the current time step from snapshots from the long-time limit. The plots show the resulting accuracy for $W/J=6.4$, $U/J=2.9$. }
\label{fig:thermalization2}
\end{figure}

In order to study the dynamics of the quantum many-body system, we here compare snapshots from the current time step to the long-time limit. In a thermalizing system, the long-time limit corresponds to a thermal equilibrium state and the scheme is thus basically the same as the thermalization learning scheme introduced in the main text. This is, however, not the case for the MBL phase.
In Fig.~\ref{fig:thermalization2}, the accuracy achieved on a test set not used during training is shown as a function of time. In each time step, the neural network parameters are optimized to enable the classification of snapshots into the categories \emph{current timestep} versus \emph{long-time limit}. This procedure has the advantage that the features used to make the classification can vary for different time steps and in particular, the network specifically searches for differences between the current time and the long-time limit. It is therefore in principle capable of identifying specific observables that have not yet reached their long-time value. 
\\
\begin{figure*}
\centering
\epsfig{file=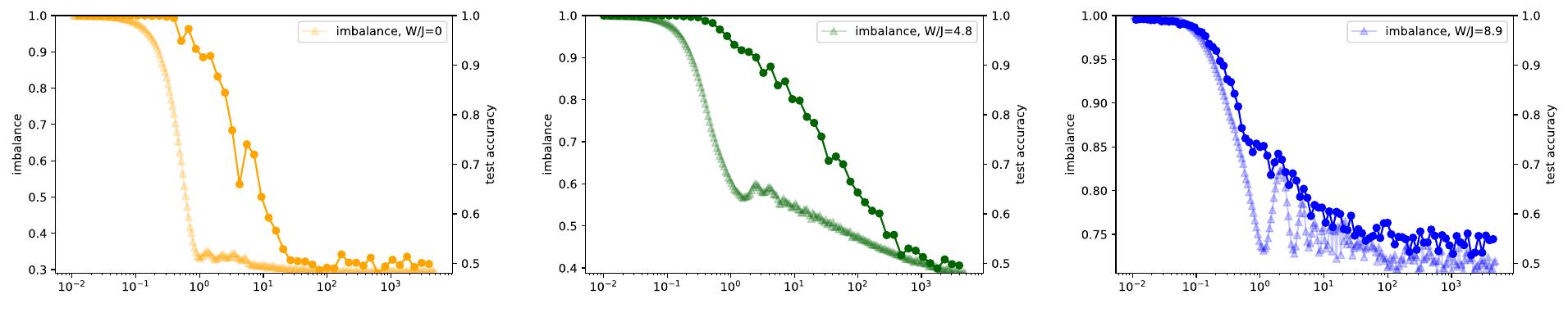, width=1\textwidth}
\caption{\textbf{Learning thermalization -- 2.} The system is initialized in a charge density wave state and the ensuing time evolution is investigated. In each time step, the neural network is trained to distinguish snapshots from the current time step from snapshots from the long-time limit. The plots show the resulting accuracy for a) $W/J=0$, b) $W/J=4.8$ and c) $W/J=8.9$. A high accuracy indicates that the current time step can be easily distinguished from the long-time limit. The imbalance is shown for comparison.} 
\label{fig:thermalization1}
\end{figure*}
In Fig.~\ref{fig:thermalization1}, the accuracy achieved on a test set not used during training is shown as a function of time when starting from the product state 
\begin{equation}
\ket{\psi_0} = \ket{2020202020}
\end{equation}
for $W/J=0,4.8,8.9$. We compare the resulting accuracy to the imbalance, defined as 
\begin{equation}
\mathcal{I} = \frac{1}{L\cdot N_s} \sum_s \sum_i \left(n_i^s - n_i^{\text{ref}}  \right),
\end{equation}
where $N_s$ is the number of snapshots,  the first sum runs over all snapshots,  $n_i^s$ is the occupation of site $i$ in snapshot $s$,  and $n_i^{\text{ref}} = 0(2)$ for $i$ even (odd). 
\\
In all three cases, the first tunneling events cause a sharp decay in the imbalance on a time scale of one hopping time. The accuracy with which the neural network can distinguish the current time step from the long-time limit is always larger than the difference of the imbalance to its long-time limit.
\\
In the many-body localized case, Fig.~\ref{fig:thermalization1}c), the accuracy shows qualitatively the same behavior as the imbalance. However, in the critical phase, Fig.~\ref{fig:thermalization1}b), there is no fast initial decay in the accuracy and instead, it is still higher than 50\%, which corresponds to its lower limit, after several hundred hopping times. Without disorder, Fig.~\ref{fig:thermalization1}a), the imbalance has almost reached its long-time value after one hopping time. The accuracy with which the network can distinguish the current time step from the long-time limit decays on a slower time-scale of about ten hopping times.


\end{document}